# Mining university rankings: Publication output and citation impact as their basis

Nicolas Robinson-Garcia[1,2]*, Daniel Torres-Salinas[3], Enrique Herrera-Viedma[4] and Domingo Docampo[5]

[1] INGENIO (CSIC-UPV), Universitat Politècnica de València, Valencia (Spain),
[2] School of Public Policy, Georgia Institute of Technology, Atlanta GA (United States),
[3] University of Granada (EC3metrics & Medialab UGR), Granada (Spain)
[4] Department of Computer Science and Artificial Intelligence, University of Granada, Granada (Spain)
[5] atlanTTic Research Center for Communications, University of Vigo, Vigo (Spain)

* Correspondence to elrobinster@gmail.com

## Abstract

World University rankings have become well-established tools that students, university managers and policy makers read and use. Each ranking claims to have a unique methodology capable of measuring the 'quality' of universities. The purpose of this paper is to analyze to which extent these different rankings measure the same phenomenon and what it is that they are measuring. For this, we selected a total of seven world-university rankings and performed a principal component analysis. After ensuring that despite their methodological differences, they all come together to a single component, we hypothesized that bibliometric indicators could explain what is being measured. Our analyses show that ranking scores from whichever of the seven league tables under study can be explained by the number of publications and citations received by the institution. We conclude by discussing policy implications and opportunities on how a nuanced and responsible use of rankings can help decision making at the institutional level

# 1. Introduction

More than a decade has gone by since the launch of the Shanghai Ranking in 2003. Nowadays, world university rankings are well-established tools that students, university managers, researchers and policy makers read and use (Hazelkorn 2008). Despite a general consensus within the bibliometric community with regard to their limitations and simplistic view of the global higher education system (Billaut et al. 2009; Rauhvargers 2014; Van Raan 2005), world university rankings have succeeded mainly because they help to navigate the complex landscape of worldwide higher education. Such is their success that international classifications have become a market niche. In fact, the main bibliometric data producers, Elsevier (Scopus) and Clarivate Analytics (Web of Science), have partnered with the main world rankings to provide the raw data for their indicators (Robinson-Garcia & Jiménez-Contreras 2017).

Each ranking employs its own methodology and indicators. For instance, the Leiden Ranking (Waltman et al. 2012) focuses only on research performance and only uses scientific production data. The Shanghai Ranking introduces as well as publication and citation data, variables related to the number of Nobel Prizes or full-time equivalent staff (Liu & Cheng 2005). The Times Higher Education World Universities Ranking includes staff-student ratios and a reputational survey within its variables (Bookstein et al. 2010). Still, all rankings render similar results, evidencing some level of agreement as to what they are measuring (Aguillo et al. 2010). In this paper, we question what university rankings are measuring, if measuring anything at all. We investigate what really lies behind the forest of university rankings, which does not allow policy makers to see what is fundamental to make decisions. Specifically, we compare the results gathered through the use of a few well-known bibliometric indicators with those offered by world university rankings and hypothesize that the weight given by all rankings to publication output and citation impact tends to overrule other methodological choices.

The paper is structured as follows. Section 2 reviews previous literature on university rankings. We divide our review in two parts: one focused on methodological aspects and a second one focused on how they are used in university management. Section 3 describes the materials and methods used in our investigation. Next, we presents the statistical results obtained after a series of principal component analyses have been carried out on the dataset. In section 5 we discuss our findings and their policy implications. Finally, we discuss our main conclusions.

## 2. Literature review

The demand for university rankings responds to two phenomena. First, the development of so-called New Public Management in higher education, especially through the 1980s and 1990s (De Boer et al. 2007) has created the demand for building quantitative tools and indicators to help policymakers and university managers to make informed decisions. Added to this phenomenon, the globalization of the higher education market has led countries and universities, not only to monitor and assess their own performance, but to compete with other higher education models and institutions worldwide (Hazelkorn 2011). Rankings producers have been quick in grasping the opportunity of fulfilling the need for analytical tools. As a result, they have positioned rankings as a comprehensive method to enable meaningful comparisons among international academic institutions and national higher education systems. At a first stage, the rise of world rankings led to discussions about methodological choices and the criteria employed at assessing universities (Van Raan 2005) derived from the difficulty of comparing such heterogeneous institutions (Collini 2012). At the same time, rankings were seen as black boxes (Florian 2007; Mutz & Daniel 2015) which needed to be deciphered (Docampo 2013) and explained (Waltman et al. 2012).

## 2.1. Methodological aspects

Bibliometricians have focused their efforts on developing advanced and nuanced methodologies to assess and benchmark the research performance of universities, but have failed in developing an intuitively, easy to interpret tool. However, simplicity does not come without a cost; hence, much of the literature regarding rankings has focused on the limitations of rankings to help in making informed decisions. Let us briefly discuss the main limitations identified in the literature. First of all, there are inherent methodological issues on the data collection process which cannot be solved through a top-down approach (Van Raan 2005; Waltman et al. 2012). Secondly, rankings fully or partially based on bibliometric data introduce a disciplinary bias which benefits certain institutions (Calero-Medina et al. 2008), a bias that cannot be fully overcome by developing rankings by fields and disciplines (Robinson-Garcia & Calero-Medina 2014). Institutional size and country location also affect the rank of institutions (Frenken et al. 2017). Most importantly, rankings consider universities as homogeneous entities which follow the same objectives and are therefore, comparable (Collini 2012). What is more, the allocation of resources by each national system influences the pertinence of the variables used to assess universities, increasing the complexity of the task (Docampo & Cram 2017).

A more pragmatic path is that adopted by a second stream of literature focused on developing a nuanced understanding on how to interpret university rankings. By acknowledging the inevitability of preventing university managers from using rankings, these studies focus their efforts on analyzing underlying factors behind the positioning of universities. In this case, studies tend to either be ranking-specific (e.g., Bornmann et al. 2013; Docampo & Cram 2015; Frenken et al. 2017; Mutz & Daniel 2015) or compare a set of rankings to identify discrepancies and common elements between them (Moed 2017a; Safón 2013). These studies have found that country differences explain in many cases the scores achieved by universities, even beyond university differences (Bornmann et al. 2013), with rankings being geographically biased towards different regions (Moed 2017a). The influence of English-speaking countries is especially noticeable in the THE Ranking (Moed 2017a). On the other hand, ARWU is strongly biased towards US universities (Safón 2013). Furthermore, institutional size, geographical characteristics such as cities' size, research orientation, or annual income, influence greatly universities' research performance in league tables (Frenken et al. 2017; Safón 2013).

A critical aspect has to do with discrepancies in scores and positioning found between rankings (Moed 2017a; Olcay & Bulu 2017) and whether these differences respond to measures of different aspects or prove the ineffectiveness of rankings. These studies point towards the central question of this study, which brings in a large set of university rankings to better understand the specific weight bibliometric indicators have on the final scores provided by rankings.

## 2.2. The use of rankings in university management

Rankings' limitations have not prevented university managers and science policy makers from employing university rankings in their daily practice. In fact, more than 90% of university managers want to improve their position in national and international rankings (Hazelkorn 2008 p. 196). They see these tools as a means to improve the public image of the institution, students' recruitment and attraction of talent. What is more, they introduce institutional policies directed at improving their position (Hazelkorn 2008). Other times, national governments may introduce formulas to improve the positioning of their institutions or use them in their political discourse as a mention of pride or critique (Salmi & Saroyan 2007). Rankings produce a reaction to the different stakeholders of the higher education system introducing external pressures which make them impossible to ignore (Sauder & Espeland 2009).

The use of rankings as a means to set strategic goals has also led to arguable practices such as the suggestion of merging institutions as a way to improve the overall position of the resulting university. Such practices are rooted in the undisputable effect size bears on ranking tables (Docampo & Cram 2015; Zitt & Filliatreau 2007). An example of this is the case of France, in which the so-called "Initiatives d'excellence" (IDEX), has pushed towards institutional mergers to increase the international visibility of the French higher education system (Docampo et al., 2015). There is a rationale for merging institutions, such as reuniting the pieces of what was formerly a single institution (as the four projects in Paris are trying to accomplish with the heirs of "Université de Paris"). Still, rankings should not, by and large, be a decisive factor, not only on the grounds that they may result in just a moderate improvement of the merging institutions, but mainly because mergers per se do not necessarily guarantee better research performance (Moed et al. 2011).

On the other hand, a responsible use of these tools can lead to positive changes and strategies at the institutional an national level (Dill & Soo 2005). World university rankings facilitate useful and enlightening comparisons among national higher education systems (Docampo 2011), while providing a solid ground to analyze different models of governance (Aghion et al. 2010) or gain a proper insight into perceptions of reputation and prestige (Moed 2017a).

## 3. Material and methods

### 3.1. Data

This study analyzes a number of world university rankings and compares them through the use of well-known bibliometric indicators of widespread use. The selection of rankings was determined based on two criteria so that the study was methodologically viable. First, rankings should encompass the global higher education landscape; hence all national rankings were discarded. Second, rankings should offer a composite indicator to classify institutions. The Leiden Ranking was therefore discarded, as it does not offer or propose an indicator by which universities should be ranked but rather offers a battery of indicators that can be used by the reader to establish comparisons. Also, the QS Top World Universities Rankings were disregarded since they do not offer complete information on all the scores universities receive. The same reasoning applied to the SCImago Institutions Rankings, which were also excluded. Table 1 includes a list of the rankings analyzed in this paper, along with a link to their website, the weight bibliometric indicators have on the final score used to rank institutions, and the database used to extract publication and citation data.

**Table 1. General overview of selected rankings based on their 2016 edition**

| Ranking | Website | Weight of Bibliometric indicators | Bibliometric data source |
|---|---|---|---|
| **Shanghai Ranking** | http://shanghairanking.com | 60% | Web of Science |
| **THE World University Rankings** | https://www.timeshighereducation.com/world-university-rankings/ | 38.5% | Scopus |
| **US News Best Global Universities** | https://www.usnews.com/education/best-global-universities/rankings | 75% | Web of Science |
| **NTU** | http://nturanking.lis.ntu.edu.tw/ | 100% | Web of Science |
| **CWUR Rankings** | http://cwur.org/ | 20% | Web of Science |
| **URAP** | http://www.urapcenter.org/2017/ | 100% | Web of Science |
| **Round University Rankings** | http://roundranking.com/ | 26% | Web of Science |

## 3.2. Description of selected rankings

Next, we briefly describe the set of rankings selected for our analysis as well as the main methodological differences observed between them. The Shanghai Ranking and the Times Higher Education World University Rankings (THE Rankings hereafter) are among the first world rankings ever established (Aguillo et al. 2010), launched in 2003 and 2004 respectively, and are widely known and used. While the former combines bibliometric indicators with Nobel prizes, Field Medals and institutional census data, the THE Rankings are widely known for combining survey data with publication, economic and institutional census data. The US News Best Global Universities league tables are more recent (2014) although they are developed by the US News & World Report, a reputed company on the elaboration of university rankings. US News & World Report is mostly known for publishing annually since 1983 the America's Best Colleges report with league tables for US higher education institutions. In this case, reputational data from surveys is also retrieved and combined along with bibliometric data. The NTU Ranking, also known as the Performance Ranking of Scientific Papers for World Universities, is developed by the National Taiwan University and released since 2007. This ranking uses uniquely bibliometric data.

CWR Rankings stands for Center for World University Rankings. This center, currently headquartered in the United Arab Emirates, launched its first edition in 2010. This ranking uses similar criteria to those employed by the Shanghai Ranking. It adds patent publication data, does not include staff data and the weight of bibliometric data is ostensibly lower. Along with the NTU Rankings, URAP (University Ranking by Academic Performance) is one of the three rankings in our selection which only employs bibliometric data. In this case, and similarly to what the Shanghai Ranking does, it employs the bibliometric suite InCites. It is produced by the URAP center, a non-profit organization and launched its first edition in 2010. It ranks more than 2500 higher education institutions worldwide. Finally, the last ranking selected is the Round University Rankings (RUR), founded in 2013 by the RUR Rankings Agency, based in Russia. This ranking uses 20 indicators and claims to measure 4 areas of universities' activities: teaching, research, international diversity and financial sustainability. It combines bibliometric data with statistical and reputational (survey) data.

## 3.3. Bibliometric indicators used in the paper

Along with data collected on the seven rankings, we used a set of bibliometric indicators that bear different degrees of dependence on the size of institutions. These were retrieved from the Clarivate's Web of Science database: the bibliometric suite InCites. This means that when establishing comparisons with results from rankings using Elsevier's Scopus as their bibliometric data source, some differences can be explained due to differences in the databases. We expect to find more similarities between bibliometric indicators and rankings based on data extracted from the Web of Science.

We use the following indicators to compare them against rankings' scores:

PUB: Total number of citable documents indexed in the Web of Science database (articles and reviews) authored by an institution in the 2011-2014 period.

H-Index: The h-index (also known as Hirsch index) was introduced by J. Hirsch in 2005. Here we calculate this indicator at the institutional level. Therefore, a university will have an h-index of *h* if it has at least *h* of its publications have received at least *h* citations.

CNCI: Category Normalized Citation Impact of the records used to compute indicator PUB. It is calculated by dividing the actual count of citing items by the expected citation rate (baseline) for publications with the same document type, year of publication and subject area. When a document is assigned to more than one subject area, an average of the ratios of the actual to expected citations is used. Therefore, we will consider that universities with a CNCI above 1 will have a higher impact than the world average, and below 1 will have a lower impact than the world average.

## 3.4. Methodological design

Scores on seven rankings from a sample of 356 universities are available for inspection and analysis. Altogether, we take the scores as seven sub-scales from a broader scale that would encompass the measures taken by the seven rankings. Conceptually, a scale is composed of several values to measure a single construct. The items within the scale should be interchangeable so that they can be taken as different ways to inquire about the same underlying characteristic that is difficult to measure directly.

We are interested in gauging the dimensionality of the scale using Principal Component Analysis (PCA), a powerful technique to extract a small set of meaningful components that summarize the correlation among a set of variables (Tabachnick & Fidell 2007). Rather than providing a theoretical explanation of the phenomenon under investigation, PCA helps on understanding the internal structure of the data, a good starting point for a further conceptual analysis.

Principal Component Analysis was used to elucidate whether the seven rankings selected were measuring the same phenomenon. That is, whether there is a single coherent variable that contributes to explain the population variance to such an extent as to be taken as a measure of a unique underlying trait that captures most of the correlations among the variables. If we were to find that one principal component accounts for a large share of the variance of the sample, it would show that the scale (composed of the seven ranking measures) is unidimensional. Therefore, the first stage of our research will be related to the number and nature of the principal components, and their importance as far as the explanation of variance is concerned.

To test the validity of a scale we should first assess its reliability, e.g., its internal consistency, and then answer the question of whether the scale represents the theoretical construct it is meant to measure. We have examined the internal consistency of the scale using standard reliability measures.

Further investigation is needed to assess the validity of the construct resulting from the interpretation of the principal components arising from PCA. We would need a different measure of the same theoretical construct in order to make the appropriate comparisons. Based on our hypothesis, - that the underlying factor is related to research performance, we use the bibliometric indicators described in Section 2.3, which we know are connected to research outcomes in terms of publications and citations. We begin by adding the H-index to our pool of seven measures, to assess the validity of the constructed scale. Finally, to refine our analysis, we introduce the other two indicators: PUB, which is size-dependent, and CNCI, which is size-independent. This will provide a better understanding of our research question.

### 3.5. Secondary analysis: The teaching component

As shown in section 4, we have not been able to identify a teaching component in our analyses. To shed some light on this issue, we did the following experiment. Using the initial scale of seven ranking measures, we removed the Times Higher Education (THE) rankings and added its Research and Teaching scores instead. The results of this analysis are included in the end of section 4.3.

## 4. Results
### 4.1. Objective of the study

Despite using different methodologies, results rendered by most world universities rankings are similar, suggesting that they are measuring, each on their own way, a similar phenomenon. However, to our knowledge, no other study has analyzed if this is the case empirically, other than showing a high correlation between the scores universities receive in each ranking (Aguillo et al. 2010). Therefore our first research question is as follows: Are all world university rankings measuring the same phenomenon, whichever that is?

If this is so, the next question to address is, what it is that rankings are measuring. For this second part, we hypothesize that, despite methodological differences as well as differences on weights assigned to publication and citation data, all of these rankings tend to measure a combination between the size of output and relative citation impact of universities. Therefore, despite claims on the accuracy and reliability each rankings make on their methodology, bibliometric indicators combining these two aspects can offer a very similar image of the higher education landscape as the one shown by university rankings.

This would have important policy implications for university managers, as it means that rankings reflect a very specific aspect of universities' performance, allowing to use these tools in an informed way, acknowledging the well-known limitations and opportunities bibliometric indicators have to offer when monitoring research performance (Hicks et al. 2015).

## 4.2. Exploratory Principal Component Analysis

Two main issues concerning the suitability of our dataset for factor analysis should first be sorted out, namely: sample size and strength of the correlation among the items (ranking scores). The length of our sample seems to be large enough. Tabachnick and Fidell (2007) have extensively reviewed this issue and concluded that "it is comforting to have at least 300 cases for factor analysis" (p. 613). To complete the analysis of the suitability of the dataset for exploratory factor analysis we need to check the correlation matrix. After inspecting the correlation structure of the seven scores (Table 2), we observe that all the correlation coefficients are highly significant ($p<0.001$, 1-tailed). Sampling adequacy was addressed through the Kaiser-Meyer-Olkin (KMO-MSA) statistic, a measure of the proportion of variance among variables that might be common variance, showing a range of values from 0 to 1. The larger the proportion of common variance, the higher the value of the KMO test and the more suited the data for factor analysis (Kaiser and Rice, 1974). The value for our sample was 0.87, showing that the correlation matrix is appropriate for factor analysis (Dziuban and Shirkley, 1974).

**Table 2 Pearson correlation matrix of selected rankings**

|       | ARWU | THE  | NTU  | CWUR | RUR  | URAP |
|-------|------|------|------|------|------|------|
| USN   | 0.87 | 0.83 | 0.83 | 0.80 | 0.73 | 0.80 |
| ARWU  | 1.00 | 0.82 | 0.90 | 0.89 | 0.70 | 0.88 |
| THE   |      | 1.00 | 0.81 | 0.70 | 0.85 | 0.77 |
| NTU   |      |      | 1.00 | 0.80 | 0.71 | 0.98 |
| CWUR  |      |      |      | 1.00 | 0.60 | 0.77 |
| RUR   |      |      |      |      | 1.00 | 0.66 |

In a well-constructed and meaningful scale, the magnitude of the correlations between its items should be relatively large, as it happens to the broad scale composed of the seven ranking scores. Reliability, a measure of the internal consistency of the data from the seven rankings in our sample, can be dealt with through the computation of Cronbach's Alpha to assess the homogeneity of the items comprising the scale (DeVellis 2003). In the sample, the Cronbach's coefficient was 0.95. We can, therefore, conclude that the scale composed of the seven ranking scores shows excellent internal consistency.

Then, we assessed the properties of the scale constructed with the seven ranking scores by examining data for 356 institutions to conduct a Principal Component Analysis. The PCA reveals the existence of just one principal component that explains over 82% of the variance of the dataset (table 3).

Loadings on this component are relevant for all variables (table 3). Loadings are related to the degree of correlation between each indicator and the component. Communalities (share of variance explained by the component) associated to the first factor (table 3), reveal that the variance of the seven rankings is adequately accounted for by the first principal component, albeit in a relatively moderate way in the case of the RUR Ranking. This indicates that the seven rankings under analysis compose a one-dimensional scale.

**Table 3 PCA Analysis results. Total variance explained and loadings of ranking indicators are only shown for Component 1**

| Component | Total Eigen Value | % Variance |
|---|---|---|
| Component 1 | 5,791 | 82,733 |

| Ranking | Loading for the component 1 | Communalities % of variance explained by the component 1 |
|---|---|---|
| USN | 0.92 | 0.85 |
| ARWU | 0.96 | 0.91 |
| THE | 0.91 | 0.82 |
| NTU | 0.95 | 0.91 |
| CWUR | 0.88 | 0.77 |
| RUR | 0.82 | 0.67 |
| URAP | 0.92 | 0.85 |

## 4.3. Hypothesis testing: Bibliometric indicators and university rankings

Given the fact that all the rankings make use of indicators related to the scientific production of academic institutions, both in quantity and quality, we hypothesize that what all the ranking are actually measuring, i.e. the factor underlying the one-dimensional character of the dataset, is related to publication output and citation impact. As a first step in our analysis of the validity of the scale composed of the seven scores, we propose the inclusion of a well-known measure that bundles output and citation impact, the institutional H-index, directly in the indicators, in a "fiducial" approach to principal components analysis (Docampo & Cram, 2014). Hence, we now have 8 variables in the dataset, the scores from the seven rankings and the H-index (computed for publications within the 2011-2015 period). All the correlation coefficients remain highly significant ($p<0.001$, 1-tailed). The KMO-MSA value is even larger, 0.89. Besides, in the new dataset of 8 variables, the Cronbach's coefficient was larger as well, 0.96; we can, therefore, conclude that the H-index shows an excellent consistency with the group of the seven ranking scores.

**Table 4 Correlation of the H-index with selected rankings.**

|  | USN | ARWU | THE | NTU | CWUR | RUR | URAP |
|---|---|---|---|---|---|---|---|
| H-Index | 0.86 | 0.91 | 0.84 | 0.96 | 0.78 | 0.75 | 0.93 |
|  | p<<.001 | p<<.001 | p<<.001 | p<<.001 | p<<.001 | p<<.001 | p<<.001 |

The results of the PCA on the new dataset reveal the existence of just one principal component, this time explaining almost 84% of the variance (table 5). The loadings on the first principal component reveal that the variance of the seven rankings continues to be adequately accounted for by the first principal component; The first component extracts 93% of the variance of the H-index (peak value), confirming our hypothesis that rankings by and large measure a combination of quantity and quality of the scientific production.

**Table 5 Total variance explained and loadings of ranking indicators for Component 1 after introducing the H-Index in the PCA Analysis**

| Component | Total Eigen Value | % Variance |
|---|---|---|
| Component 1 | 6,71 | 83.8% |

| Ranking | Loading for the component 1 | Communalities % of variance explained by the component 1 |
|---|---|---|
| USN | 0.92 | 0.84 |
| ARWU | 0.96 | 0.91 |
| THE | 0.90 | 0.82 |
| NTU | 0.96 | 0.92 |
| CWUR | 0.87 | 0.76 |
| RUR | 0.82 | 0.67 |
| URAP | 0.93 | 0.87 |
| **H-Index** | **0.96** | **0.93** |

To complete the analysis, we try to discern the influence of institutional size on the positioning of rankings. For this, we now introduce two indicators that point into or away from that direction: Number of articles of a university recorded in the web of Science (PUB), a size-dependent measure; and Average number of citations of the publications of a university, normalized for field and publication year (CNCI), a size-independent measure. We add those results to our dataset, which is now composed by 10 variables. The correlation matrix of the indicators with rankings is shown in table 6. The KMO-MSA value is now 0.9, while, the Cronbach's coefficient is 0.96. We can therefore conclude that the both PUB and CNCI show an excellent consistency with the seven ranking scores and the H-index.

**Table 6 Correlation matrix results including bibliometric indicators**

|  | USN | ARWU | THE | NTU | CWUR | RUR | URAP | H-Index | Nr Pubs | MNCS |
|---|---|---|---|---|---|---|---|---|---|---|
| H-Index | 0.86 | 0.91 | 0.84 | 0.96 | 0.78 | 0.75 | 0.93 | **1.00** | 0.88 | 0.69 |
| PUB | 0.72 | 0.83 | 0.68 | 0.93 | 0.71 | 0.61 | 0.96 |  | **1.00** | 0.35 |
| CNCI | 0.71 | 0.63 | 0.77 | 0.59 | 0.55 | 0.61 | 0.50 |  |  | **1.00** |

The PCA analysis reveals the existence of two principal components, explaining over 88% of the variance (table 7). The first, dominant component is associated with the size-dependent bibliometric indicator chosen, PUB, whereas the second component appears to be linked to the size-independent bibliometric indicator chosen, CNCI (Figure 1). Table 7 shows that after varimax rotation the first component explains around 52% of the variance, while the second explains around 36% of the variance. Shares of variance for each variable explained by the first and second principal components are shown in table 8. The effect of the size-independent component is higher for those international classifications that make use of at least one indicator related to the average number of citations per paper from an institution.

**Table 7 Results of PCA Analysis including all bibliometric indicators.**

|  | Initial eigenvalues | | | Varimax Rotation | | |
| --- | --- | --- | --- | --- | --- | --- |
|  | Total | % of Var. | Cum. % | Total | % of Var. | Cum. % |
| Component 1 | 7.95 | 79.54 | 79.54 | 5.214 | 52.138 | 52.138 |
| Component 2 | 0.90 | 9.02 | 88.55 | 3.642 | 36.416 | 88.554 |

**Figure 1 Scatterplot of PCA Analysis results. Cases in black and circle are university rankings. Cases in red denote bibliometric indicators.**

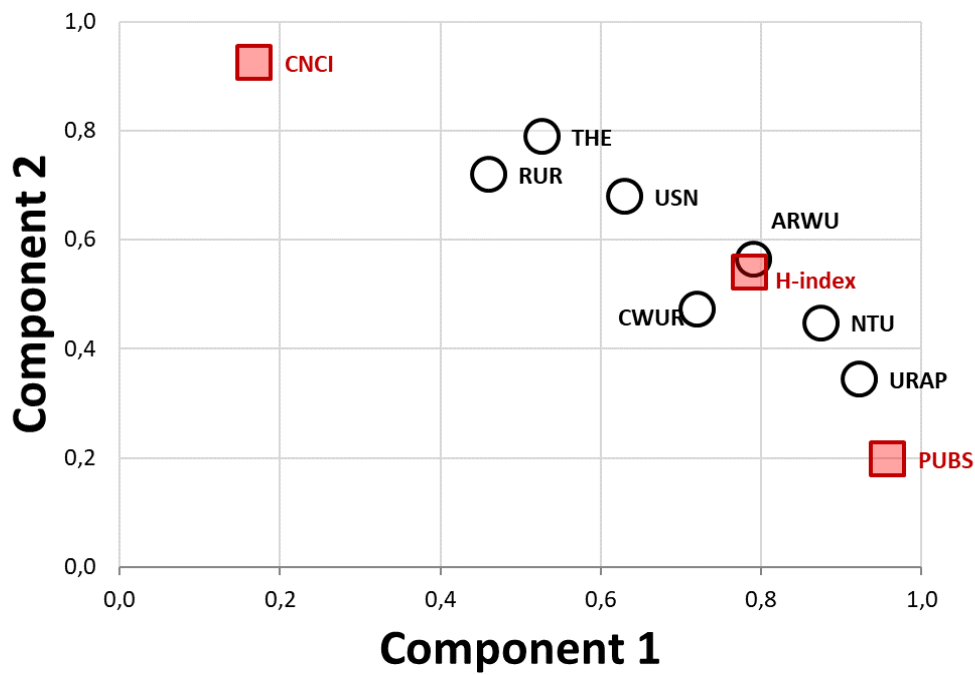

**Table 8 Total variance explained for the two components identified after introducing CNCI and PUB in the PCA model**

|  | Loading for the component 1 and 2 | | | Communalities % of variance explained by the component 1 and 2 | |
|---|---|---|---|---|---|
|  | Component 1 | Component 2 | TOTAL | Component 1 | Component 2 |
| □ PUB | 0.92 | 0.04 | 0.96 | 96% | 4% |
| ○ URAP | 0.85 | 0.12 | 0.97 | 88% | 12% |
| ○ NTU | 0.77 | 0.20 | 0.97 | 79% | 21% |
| ○ CWUR | 0.52 | 0.22 | 0.74 | 70% | 30% |
| ○ ARWU | 0.62 | 0.29 | 0.91 | 68% | 32% |
| □ H-Index | 0.63 | 0.32 | 0.94 | 66% | 34% |
| ○ USN | 0.40 | 0.46 | 0.86 | 46% | 54% |
| ○ THE | 0.28 | 0.62 | 0.90 | 31% | 69% |
| ○ RUR | 0.21 | 0.52 | 0.73 | 29% | 71% |
| □ CNCI | 0.03 | 0.85 | 0.88 | 3% | 97% |

□ Bibliometric Indicators
○ Ranking

An interest finding is the lack of a teaching component in our analyses. Figure 2 shows the result of splitting the THE ranking (the only one with a teaching score) into two rankings: research and teaching scores. By forcing two principal components we observe that the first principal component still explains over 80% of the variance in the sample. Two Principal components extract almost 90% of the variance. In spite of that, it is not possible to tell apart the two indicators from the ranking THE, showing that teaching and research reputation go together. Note that the x-axis starts on 0.8 due to visualization issues.

**Figure 2 Scatterplot of PCA Analysis results differentiating THE research and teaching scores. Only university rankings are included**

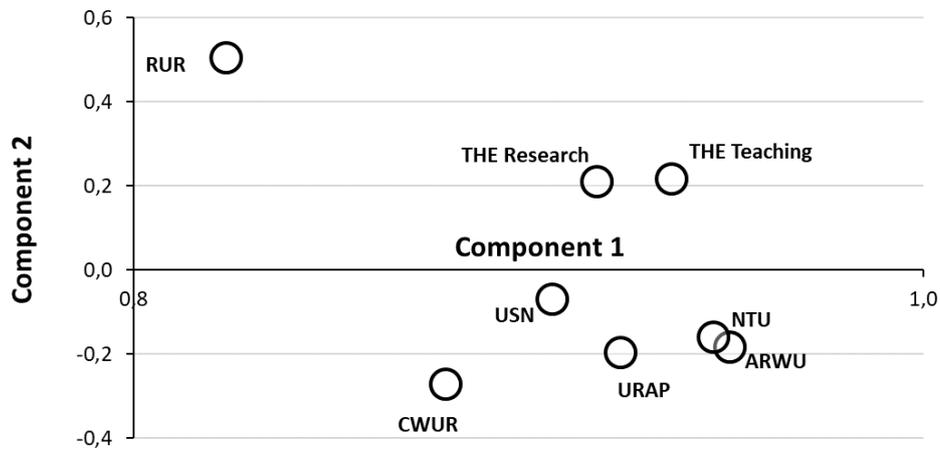

## 5. Discussion and policy implications

We began the paper by acknowledging the currency that international academic classifications are gaining nowadays. We have described a number of global rankings based on different methodologies and indicators. In spite of differences in methodologies, it is commonly agreed that rankings provide compatible results. We therefore posed a legitimate research question. One that, − if solved, − would shed light on what university rankings are measuring, if measuring something at all.

**Figure 3 Scatterplot of universities according to their score in the Shanghai Ranking and their institutional H-Index**

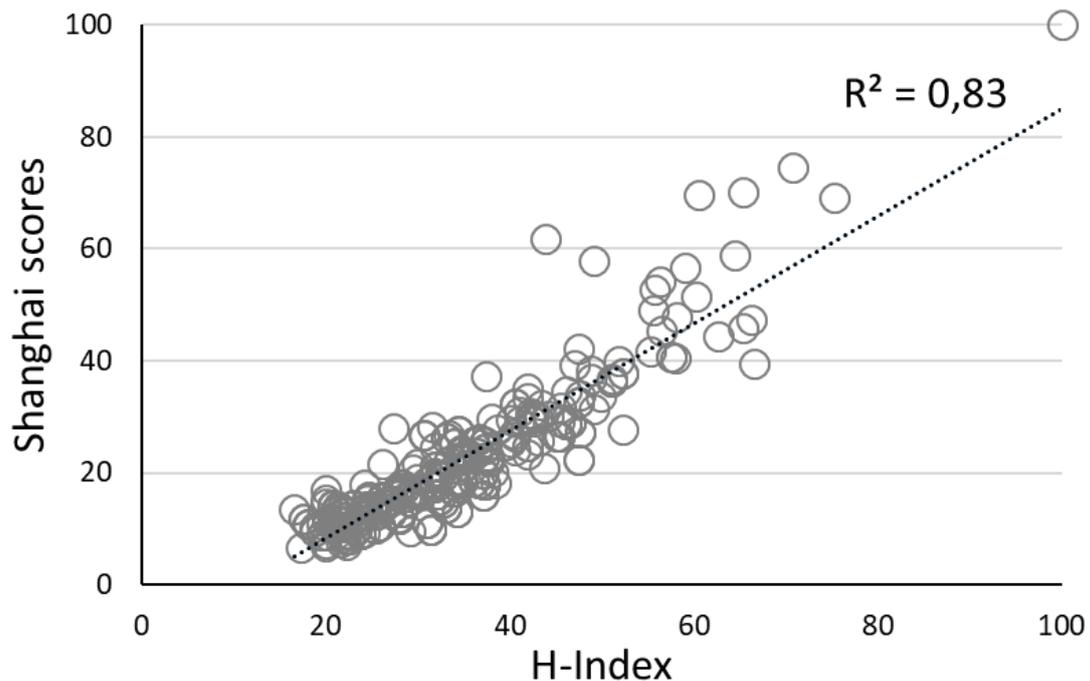

We selected an empirical framework for our correlational study, conceptualizing the dataset of measures (scores) from the seven international rankings selected as the results of a scale of which we want to find out whether it is reliable and valid. Using Principal Component Analysis we have been able to elucidate our research question; we have found that the scale is one-dimensional, and we have also found that what rankings measure is in one way or another related to counts of publications and citations. We have been able to articulate our answers by using straightforward bibliometric indicators that helped us in explaining how rankings end up measuring a compact of size-dependent and independent research variables in a coherent way.

There are a number of different features, including teaching related measures, that rankings include in their portfolio of indicators. What we maintain here is not that those features are not measured, but rather that their influence in the final score becomes highly overshadowed by the bibliometric information, no matter the weights used on the different sets of indicators. We were particularly puzzled by the fact that teaching indicators would not be identified as characteristics related to a different underlying factor. Our answer to that conundrum is that teaching scores by and large rely on surveys and our hunch is that academics respond indistinctly to surveys on teaching or research. This was further confirmed by performing additional experiments (Figure 2). Consequently, our research shows that although the rankings apparently use different evaluation criteria, publications and citations are the basis for all of them.

Our analysis also confirms that size matters when explaining institutional league tables, a well-known issue which presents important limitations when using rankings in university management (Docampo & Cram 2015). This also implies that all limitations attributed to bibliometric indicators and bibliometric databases (i.e., disciplinary biases, Mathew effect) will also be present in university rankings. Furthermore, the influence of size brings in associated factors, such as annual income or reputation, as noted elsewhere (Safón 2013). Still, the empirical demonstration of these issues does not necessarily diminish the potential usefulness of university rankings for policy makers. The high relationship between the H-Index and ranking scores can help university managers to better interpret factors explaining their institutions positioning. By doing so, the university manager can use rankings in an informed way and therefore benefit from the tool in combination with their own judgment in what is usually referred to as 'informed peer review' (Moed 2007). For instance, one could plot universities based on their ranking scores and their institutional H-Index as observed in figure 3 as a means to analyze if their institution is 'living above or under its expectations' based on their position with respect to the tendency line. Depending on the location of the university under observation, university managers would be able to look in the different indicators employed by the rankings on factors that could differences between their expected and observed position. Furthermore, these types of analyze can reveal methodological inconsistencies of the ranking itself (e.g., Full Time Equivalent counts for specific universities in the Shanghai Ranking (Docampo 2013), or institutional gaming (Moed 2017a).

## 6. Concluding remarks

University rankings, despite concerns on their validity, have now become highly regarded tools by university managers (Leydesdorff et al. 2016), and the bibliometric community should not disregard their large expansion. Hence, the key importance of developing methods, techniques and recommendations of their use, to ensure that they are employed in an informed and intelligent way. Simple analyses such as the one deployed in figure 3 can greatly enrich ranking consumers' experience and help them take a more critical look at the information provided by these tools (Moed 2017b p. 149).

While it is not the goal of this paper nor we encourage the use of university rankings for decision-making, we do acknowledge that they are important sources of information which have the potential to inform university managers when used intelligently. Hood (2012) discusses that two opposed perspectives generally emerge in public management: that which relies entirely on numbers, and second one which rejects them entirely. While the former positioning can lead, and has led, as reviewed earlier, to ill-informed decisions in university management, an informed and critical use of university rankings can improve decision-making and ease university managers' work.

## Acknowledgments

The authors would like to thank two anonymous reviewers for their comments and suggestions. Nicolas Robinson-Garcia acknowledges financial support from a Juan de la Cierva-Incorporación grant from the Spanish Ministry of Science, Innovation and Universities.